# Electron-phonon interaction during optically induced ultrafast magnetization dynamics of Au/GdFeCo bilayers

Richard. B. Wilson[1,2,3,*], Charles-Henri Lambert[1], Jon Gorchon[1,2], Yang Yang[4], Sayeef Salahuddin[1,2], Jeffrey Bokor[1,2].

1) Department of Electrical Engineering and Computer Sciences, University of California, Berkeley, CA 94720, USA
2) Lawrence Berkeley National Laboratory, 1 Cyclotron Road, Berkeley, CA 94720, USA
3) Department of Mechanical Engineering and Materials Science & Engineering Program, University of California, Riverside, CA 92521, USA
4) Department of Materials Science and Engineering, University of California, Berkeley, CA 94720, USA

correspondence should be addressed to rwilson@engr.ucr.edu and jbokor@berkeley.edu.

## Abstract

The temperature evolution of GdFeCo electrons following optical heating plays a key role in all optical switching of GdFeCo and is primarily governed by the strength of coupling between electrons and phonons. Typically, the strength of electron-phonon coupling in a metal is deduced by monitoring changes in reflectance following optical heating and then analyzing the transient reflectance with a simple two-temperature thermal model. In a magnetic metal, the change in reflectance cannot be assumed to depend only the electron and phonon temperatures because a metal's reflectance also depends on the magnetization. To deduce the electron-phonon coupling constant in GdFeCo, we analyze thermal transport in Au/GdFeCo bilayers following optical heating of the GdFeCo electrons. We use the reflectance of the Au layer to monitor the temperature evolution of the Au phonons. By interpreting the response of the bilayer to heating with a thermal model, we determine the electron-phonon coupling constant in GdFeCo to be $6 \times 10^{17}$ W m$^{-3}$ K$^{-1}$ ± 40%, corresponding to an electron-phonon relaxation time in GdFeCo of ~150 fs.

**Introduction**

Picosecond optical heating of electrons in the ferrimagnetic metal Gd$_x$(Fe$_{90}$Co$_{10}$)$_{1-x}$ can cause the direction of the metal's magnetic moment to switch [1]. This phenomena, commonly referred to as all optical switching, is induced by a nonequilibrium between the electronic, lattice, and magnetic degrees of freedom[1-4]. As the hot electrons transfer energy to the magnetic degrees of freedom of the FeCo and Gd sublattices, angular momentum is transferred between the FeCo and Gd spins [3,5-7]. If sufficient energy is transferred from the hot electrons to the magnetic sublattices over a sufficiently short time scale [4,8,9], the magnetization direction of both the FeCo and Gd magnetic sublattice switch direction [1-11].

Numerous phenomenological models have been proposed to describe the magnetization dynamics of ferrimagnetic systems following electron heating. Atomistic calculations model the magnetization dynamics of the transition metal and rare earth spins using the Landau-Lifshitz-Gilbert equation with Langevin dynamics. Alternatively, several microscopic models can successfully predict magnetization switching by considering scattering rates that govern energy and angular momentum exchange between electrons, rare earth spins, and transition metal spins.

The transient temperature response of the optically excited electrons is critical to models of all optical switching. For example, Kalashnikova and Kozub state that "reversal depends on a delicate interplay between demagnetization time and cooling time of the mobile electrons" [12]. The transient temperature response of electrons to optical heating is typically described with a two temperature model [1,3,9]. In a two temperature model, the electron temperature is a function of the electronic heat capacity, the rate of energy transfer into the electrons via the laser, and the rate that energy is transferred from the electrons to the lattice via electron-phonon coupling. However, for GdFeCo, the electron-phonon coupling constant has not been experimentally measured. As a result, theoretical studies of all optical switching instead use values for $g_{ep}$ that are "typical" of transition metals [13]. Unfortunately, typical values for electron-phonon coupling constants in transition metals vary by several orders of magnitude [14,15]. Atxitia *et al.* recently demonstrated that, at least for predictions based on atomistic simulations using the Landau Lifshitz Gilbert equation, the minimum laser energy needed to reverse the magnetization of GdFeCo varies widely depending on the value assumed for the electron-phonon coupling constant [13].

The standard method for measuring the electron-phonon coupling constant in a metal is time domain thermoreflectance [14,16]. In a standard time domain thermoreflectance measurement, the metal's reflectance is used as a thermometer. The change in the metal's reflectance $\Delta R$ is assumed proportional to the change in electron and phonon temperatures

$$\Delta R = aT_e + bT_p, \qquad (1)$$

where $T_e$ and $T_p$ are the transient temperature changes of the electrons and phonons as predicted by the two temperature model [16]. The data is fit with a two temperature model that treats $g_{ep}$ and the ratio of the proportionality constatns $a/b$ as fitting parameters. The accuracy of the value for $g_{ep}$ derived in this manner is contingent on Eq. 1 being an accurate description of the heating induced changes in the metal's reflectance.

In magnetic metals, $\Delta R$ cannot be assumed to be a reliable measure of the electron and phonon temperatures because changes to the magnetization, $\Delta M$, can also impact the reflectance. In general, any change to electron interband and intraband scattering rates at optical frequencies will modify a metal's reflectance [17]. Magnetostriction will increase the plasma frequency and cause shifts in the electron energy bands in a manner similar to thermal expansion [17]. Additionally, like thermal expansion, magnetostriction can cause shear strains within the substrate that may split degenerate energy bands in the metal [17]. An increased population of magnetic excitations such as magnons and spin fluctuations can decrease electron relaxation times via increased scattering rates [18]. Furthermore, the band structure of magnetic metals is dependent on the magnetization via exchange splitting [19]. In some magnetic metals, such as FePt:Cu, $\Delta R$ is almost entirely dominated by $\Delta M$, as evidenced by the similarity between magneto-optic Kerr effect measurements of $\Delta M$ and time domain thermoreflectance measurements of $\Delta R$ [20].

In order to measure $g_{ep}$ of GdFeCo, an accurate optical thermometer is required. We use the reflectance of an optically thick Au layer adjacent to the GdFeCo in order to probe the thermal response of the bilayer to optical heating. When an ultrafast optical pulse is absorbed by a thin GdFeCo film in a GdFeCo/Au bilayer, energy is initially deposited in the GdFeCo electrons. In the picoseconds following laser irradiation, the GdFeCo electrons transfer energy to the Au electrons, the GdFeCo phonons, and the Au phonons. By monitoring the temperature rise of the Au phonons via changes in the Au reflectance, and comparing to the predictions of a thermal model for transport in the bilayer, we determine $g_{ep,\text{GFC}} \approx 6 \pm 2.4 \times 10^{17}$ W m$^{-1}$ K$^{-1}$ in Gd$_{29}$(Fe$_{90}$Co$_{10}$)$_{71}$. Atxitia *et* *al.* predicted that for an electron-phonon coupling constant of 6 x $10^{17}$ W $m^{-3}$ $K^{-1}$, the energy density required for switching GdFeCo with a 50 femtosecond laser pulse is ~ 0.25 GJ $m^{-3}$. This compares favorably with our measurements of 0.35 $\pm$ 0.07 GJ $m^{-3}$ for the switching threshold in GdFeCo with a 55 femotsecond laser pulse [4].

**Experimental Details**

Pump/probe experiments were carried out on two $Gd_x(Fe_{90}Co_{10})_{1-x}$/Au bilayer samples deposited via magnetron sputtering on sapphire substrates. The pump/probe experiments were carried out with a Coherent Mantis oscillator with an 80 MHz repetition rate at 830 nm center wavelength. The pulse durations of the pump and probe beams were 2.3 and 0.3 ps, respectively. The pump beam was modulated with a 50% duty cycle at a frequency of 1 MHz and focused on the GdFeCo film at normal incidence with a 10x objective through the sapphire substrate. The reflected intensity of the probe light at normal incidence on the Au side of the bilayer was monitored with a Si photodiode. Pump induced intensity variations in the reflected probe beam at the 1 MHz modulation frequency were measured using an rf-lockin amplifier. Optical delay between the pump and probe beams was controlled with a linear delay stage on the pump beam path.

The $Gd_x(FeCo)_{1-x}$ was grown via co-sputtering of a Gd and $Fe_{90}Co_{10}$ target with an estimated composition of x = 0.29. In order to obtain an estimate of the composition, both targets were first individually calibrated via XRR to extract their respective deposition rate. Then the final GdFeCo atomic composition was extrapolated calculating the theoretical ratio between the different atoms. The Au film thickness for the two bilayer samples were 73 and 91 nm. The GdFeCo thicknesses, including a 2 nm Ta seed layer, were 11.5 nm. The GdFeCo and Au layer thicknesses were measured via XRR following deposition. We estimate the uncertainty in the GdFeCo thickness of 5% based on XRR measurements of half a dozen GdFeCo films deposited over several months. Ellipsometry measurements were performed on a 20 nm thick GdFeCo film capped with 2 nm of Ta in order to determine optical constants of n = 3.2 and k = 3.5. Optical constants for Au were set based on prior ellipsometry measurements of sputtered Au films [21]. Four point probe measurements of the 91 nm thick Au layer and 20 nm thick Ta capped GdFeCo layer yielded electrical resistivities for the sputtered Au and GdFeCo layers of 3 and 150 µΩ cm.

To model the thermal transport in the metallic bilayer, we used a numerical solution to coupled heat-diffusion equations [20]. In the Au layer, two coupled heat diffusion equations were solved with temperatures for the electrons and phonons, $T_{e,\text{Au}}$ and $T_{p,\text{Au}}$. In the GdFeCo, we used three coupled heat diffusion equations with three temperatures to track the thermal energy stored and transported by the electrons, spins, and phonons. We do not consider the spin temperature in the GdFeCo to be a valid descriptor of the thermodynamic state of the spin system. The transient magnetic states that can occur following laser irradiation do not occur in the equilibrium phase diagram of GdFeCo [6], and therefore cannot be described with an effective temperature. Therefore, the sole purpose of the spin temperature in our model is to account for the impact of energy transfer between the electrons and magnetic sublattices on the transient temperature response of the electrons.

All thermal properties in the model except for the electron-phonon coupling constant in GdFeCo were fixed. The electron-phonon coupling constant of GdFeCo $g_{ep,\text{GFC}}$ was treated as a fitting parameter. The electrical thermal conductivities in the GdFeCo and Au layers, $\Lambda_{e,\text{GFC}}$ and $\Lambda_{e,\text{Au}}$, were calculated to be 250 and 5 W m$^{-1}$ K$^{-1}$ from the Wiedemann-Franz law. The phonon thermal conductivity of Au was set to 3 W m$^{-1}$ K$^{-1}$ based on an extrapolation from low temperature measurements [22,23]. The phonon thermal conductivity of GdFeCo was set to 2 W m$^{-1}$ K$^{-1}$, consistent with molecular dynamics simulations for magnetic metals with comparable alloy concentrations [24]. The electron phonon coupling constant of Au $g_{ep,\text{Au}}$ was set to 2.2 W m$^{-3}$ K$^{-1}$ based on prior literature measurements [14,25]. The electronic heat capacity of Au and GdFeCo were set equal to $C_{e,\text{Au}}$ = 0.02 J cm$^{-3}$ K$^{-1}$ [25] and $C_{e,\text{GFC}}$ = 0.08 J cm$^{-3}$ K$^{-1}$ [26]. The spin and phonon heat capacities of GdFeCo were set to 0.7 and 2.25 J cm$^{-3}$ K$^{-1}$ [27], while the phonon heat capacity of Au was set to 2.5 J cm$^{-3}$ K$^{-1}$ [25]. The electron-spin coupling constant was fixed to 10$^{17}$ W m$^{-3}$ K$^{-1}$ based on measurements of an FePt:Cu alloy [20]. The electronic interface conductance between Au and GdFeCo set to 8 GW m$^{-2}$ K$^{-1}$, based on specific electrical resistance measurements of Co/Cu [28] and the interfacial form of the Wiedemann Franz law [29]. The remaining parameters in the thermal model had no impact on the temperature evolution on short time-scales, and therefore do not impact the derived value of the electron-phonon coupling constant of GdFeCo. The phonon-phonon interface conductance between the Au and GdFeCo layers and the GdFeCo and sapphire were both set to 200 MW m$^{-2}$ K$^{-1}$, typical values for phonon interface conductances [30]. The sapphire heat capacity and thermal conductivity were set to 3.1 J cm$^{-3}$ K$^{-1}$ and 30 W m$^{-1}$ K$^{-1}$ [25].

To model the light-metal interaction, we performed multilayer reflectivity calculations to calculate the distribution of absorbed laser energy as well as the depth dependence of the thermoreflectance of the Au layer [25]. The results of these calculations for the 11.5 nm GdFeCo / 73 nm Au bilayer are shown in Fig. 1.

In order to fit the predictions of our model to the data, it is necessary to know

what the proportionality constants in Eq. 1 are for Au. At an optical wavelength of 800

nm, $\Delta R$ of Au is primarily determined by the transient changes in the temperature of the Au phonons. Typically, $\Delta R$ of a normal metal is also sensitive to $T_e$, because an increase in $T_e$ broadens the step in the Fermi distribution. A broader step in the Fermi distribution affects interband transitions that originate or terminate on states near the Fermi level. However, in Au, no interband transitions occur for optical wavelengths in the near infrared [17]. As a result, $\Delta R$ of Au is primarily due to the effect of increased phonon populations on intraband scattering [17,21,31]. In our modelling, we set $b/a \approx 50$ based on previously published measurements of picosecond thermal transport in Au/Pt bilayers [25]. Alternatively, a theoretical estimate of $b/a \approx 100$ is possible by considering the temperature dependence of the electron-electron and electron-phonon intraband scattering rates at optical frequencies [31]. Increasing the value of $b/a$ in our model from 50 to 100 decreases our best fit value for the electron-phonon coupling constant of GdFeCo by ~25%.

**Results and Discussion**

To understand why the Au phonon temperature at short delay times is sensitive to the electron-phonon coupling constant in the adjacent GdFeCo film, it is useful to consider the time-scale for the Au phonons to equilibriate with the rest of the bilayer, $\tau_{Au}$. The equilibriation time between two thermal reservoirs is

$$\tau = \left( \frac{g}{C_1} + \frac{g}{C_2} \right)^{-1} \quad (2)$$

where $g$ is the coupling constant between the reservoirs, and $C_1$ and $C_2$ are the heat capacities of the two reservoirs [22]. For simplicity, we consider the case where the thermal resistance between the Au electrons and GdFeCo electrons is negligible, e.g. the Au and GFC electrons are in perfect thermal contact. In the limit that $g_{ep,\text{GFC}} \to 0$, $\tau_{Au} \sim C_{e,Au}/g_{ep,Au}$. This is on the order of 1 picosecond and the temperature of the Au phonons will easily reach its maximum value in a few picoseconds following the pump laser pulse. In the opposite limit of $g_{ep,\text{GFC}} \to \infty$, $\tau \sim C_{p,\text{GFC}}/g_{ep,Au}$. This is on the order of one hundred picoseconds [25] and the transient temperature rise of the Au phonons for the first few picoseconds following laser irradiation is negligible compared to the temperature hundreds of picoseconds later. For a value of $g_{ep,\text{GFC}}$ between these two limits, the Au phonons will heat on both these time-scales, although at different rates.

As a result, the ratio of the Au temperature for the first few picoseconds to its maximum value several hundred picoseconds later is determined by $g_{ep,\text{GFC}}$. The ratio will also be impacted by the thermal resistance between the GFC and Au electrons, and the ratio of laser energy absorbed by the Au vs. GFC electrons. All of these factors are accounted for in our thermal model.

The temperatures predicted by our thermal model are shown in Fig. 2a. Using the thermoreflectance as a function of depth shown in Fig. 1, we derive a best fit to our experimental data for the electron phonon coupling constant of GdFeCo of 6 x 10$^{17}$ W m$^{-3}$ K$^{-1}$, see Fig. 2b. To convert the experimental transient reflectance data into a temperature, we scaled the data so that the temperature rise at 300 ps is 2.5 and 2.1 K for the 73 and 91 nm thick Au layer samples, respectively. These temperatures correspond to the maximum phonon temperature that occur in the Au films for the absorbed pump fluence of ~0.5 J m$^{-2}$.

In order to estimate the uncertainty in the derived value of the electron-phonon coupling constant, we used our thermal model to determine the sensitivity of our measurement is to various thermal parameters. We define the sensitivity of the phonon temperature $T_{p,\text{Au}}$ to a thermal parameter $\alpha$ as [32]

$$S_\alpha = \frac{\partial \ln(T_{p,\text{Au}})}{\partial \ln \alpha} \qquad (3)$$

The sensitivity of our measured signal to the electron-phonon coupling constant of GdFeCo is shown in Fig. 3a. A sensitivity of -0.2 at a delay time of 5 picoseconds indicates that a 5% decrease in the electron-phonon coupling constant will produce a 1% increase in our model's prediction for $T_{p,\text{Au}}$. Also included in Fig. 3a are the sensitivities of the other thermal parameters that significantly impact the temperature of the Au phonons for the first few picoseconds. The largest uncertainty in our measurement is due to our 5% uncertainty in the GdFeCo thickness, which causes a 25% uncertainty in our derived value for $g_{ep,\text{GFC}}$. Including our uncertainty in $b/a$ (discussed above), an uncertainty of 10% in $\Lambda_{e,\text{GFC}}$ and $\Lambda_{e,\text{Au}}$, an uncertainty of 5% in the Au thickness $h_{\text{Au}}$, and an uncertainty of 25% in the electron phonon coupling constant of Au yields a total uncertainty in our derived value for the electron-phonon coupling constant of GdFeCo of 40%.

A key assumption in our analysis is that the measured changes in probe beam reflectance are not sensitive to the temperature of the GdFeCo layer. In Fig. 3b, we show calculations of the reflectance from the Au/GdFeCo bilayer as a function of Au thickness. For thicknesses larger than 70 nm, the GdFeCo contributes negligibly to the reflectance of the metallic bilayer. Therefore, we assume measurements of $\Delta R$ are not sensitive to the temperature of the GdFeCo layer when the probe beam is incident on the Au surface. To confirm this, we performed measurements on both 73 and 91 nm thick Au layers. The thin GdFeCo film underneath the 73 and 91 nm Au layer causes a 0.4 and 0.1 % deviation from the reflectance of an infinitely thick Au layer. Therefore, if the temperature evolution of the GdFeCo layer contributed to the measured $\Delta R$ in a non-negligible way, we would not be able to fit the data for both bilayers with the same set of model parameters. Measurements of both samples yielded identical results, see Fig. 2b, confirming that our measurements were not sensitive to the temperature evolution of the GdFeCo layer.

Our derived value for the electron-phonon coupling constant in GdFeCo is reasonably consistent with prior experimental or theoretical estimates of $g_{ep}$ in transistion metal magnets. For example, theoretical calculations for Ni, Fe, and Co estimate $g_{ep}$ ~10, 5, and 35 x $10^{17}$ W $m^{-3}$ $K^{-1}$, respectively [13,15,33]. An experimental estimate based on optical damage thresholds in Ni suggests $g_{ep}$ ~ 4 x $10^{17}$ W $m^{-3}$ $K^{-1}$ in Ni [14].

Our results compare favorably with the predictions of Atxitia *et al.* for the relationship between the electron-phonon coupling constant and the energy density required to switch GdFeCo. Atxitia *et al.* predicted that for an electron-phonon coupling constant of 6 ± 2.4 x $10^{17}$ W $m^{-3}$ $K^{-1}$ an energy density of 0.2 ± 0.8 J $m^{-3}$ should be necessary to switch the magnetization. In other work [4], we report that an absorbed fluence of 7 ± 1.4 J $m^{-2}$ is necessary to optically switch a 20 nm thick GdFeCo film with a 55 femtosecond laser pulse, corresponding to an energy density of 0.35 ± 0.7 J $m^{-3}$. Therefore, the agreement between our measurements and Atxitia *et al.* is quite good considering both experimental uncertainties and the number of free parameters in the atomistic simulations, e.g. coupling between the spins and the electronic bath.

Therefore, our results provide further experimental evidence that, despite their phenomenological nature, atomistic simulations based on the LLG equation are useful tools for modelling magnetization dynamics of ferrimagnets that are capable of quantitatively accurate predictions. Whether atomistic simulations based on the LLG equation are capable of replicating experimental observations of all optical switching of GdFeCo with optical pulses as long as 10 ps remains an open question [4].

In conclusion, we characterized thermal transport in two GdFeCo/Au bilayers following optical irradiation of the GdFeCo with a 2 picosecond laser pulse. During the optical irradiation of the GdFeCo, electronic heat currents carry the energy absorbed by the GdFeCo electrons throughout the bilayer. In the picoseconds following laser irradiation, strong electron-phonon coupling in the GdFeCo layer thermalizes the GdFeCo electrons and phonons. By monitoring the temperature evolution of the Au phonons via changes in optical reflectivity of the Au, we determined the electron-phonon coupling constant in the GdFeCo layer to be $g_{ep,\text{GFC}} \approx 6 \pm 2.4 \times 10^{17} \text{ W m}^{-1} \text{ K}^{-1}$. Future work will examine the ability of pure electronic heat currents generated by optical irradiation of the Au layer to switch the magnetization direction of the GdFeCo without laser energy being directly absorbed by the GdFeCo.

**Acknowledgements**

This work was primarily supported by the Director, Office of Science, Office of Basic Energy Sciences, Materials Sciences and Engineering Division, of the U.S. Department of Energy under Contract No. DE-AC02-05-CH11231 within the Nonequilibrium Magnetic Materials Program (MSMAG). We also acknowledge the National Science Foundation Center for Energy Efficient Electronics Science for providing most of the experimental equipment and financially supporting Y.Y. during his work on this project.

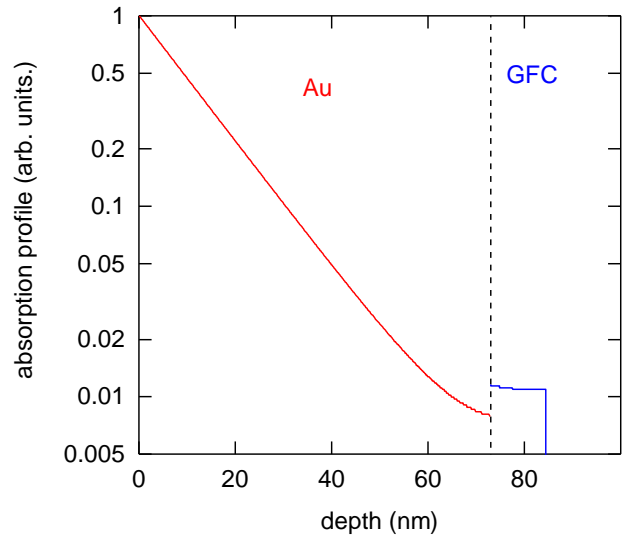

**Figure 1.** Distribution of light absorption as a function of depth in the 73 nm Au / 11.5 nm GdFeCo bilayer. The refractive indices used in the calculation were $3.2 + 3.5i$ and $0.2 + 4.9i$ for GdFeCo and Au, respectively.

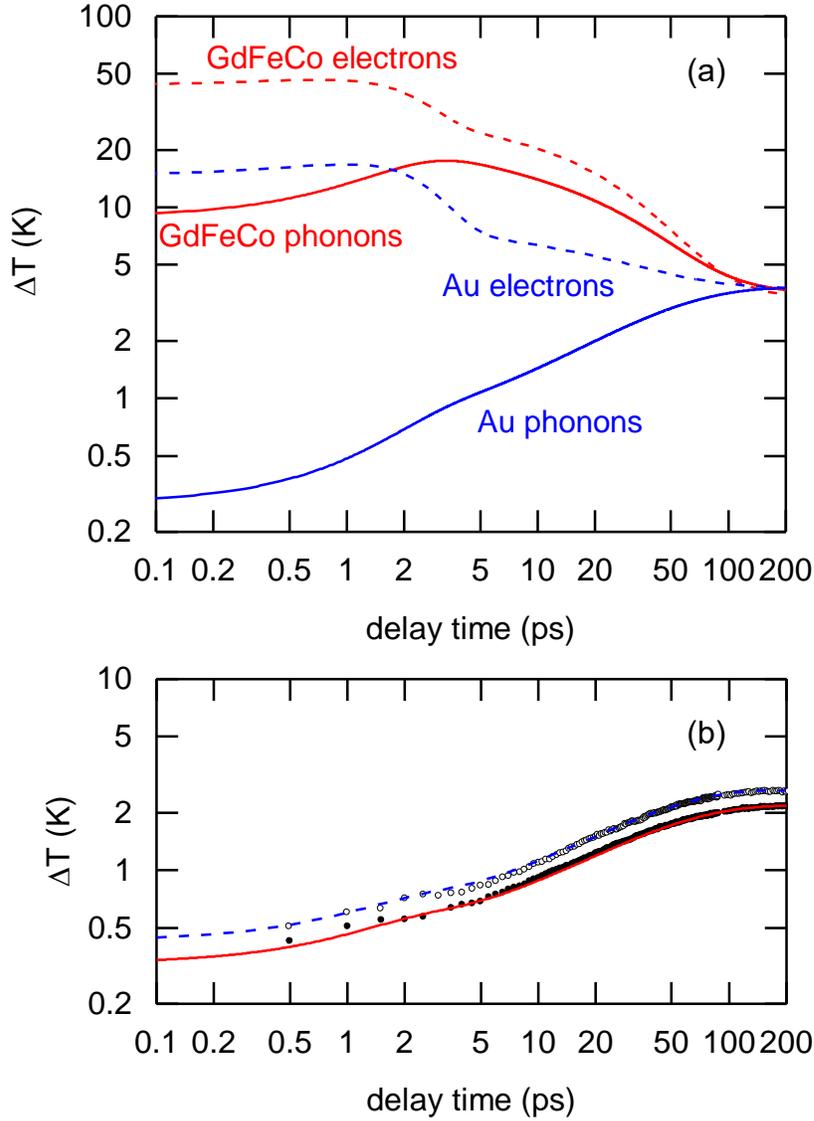

**Figure 2.** (a) Electron and phonon temperature excursions of the 73 nm Au / 11.5 nm GdFeCo bilayer calculated from the thermal model with $g_{ep,GFC} = 6\times 10^{17}$ W m$^{-3}$ K$^{-1}$. Solid lines are the phonon temperatures, while dashed lines are the electron temperatures. (b) Time-domain thermoreflectance data of the 91 nm Au / 11.5 nm GdFeCo bilayer (filled markers) and 73 nm Au /11.5 nm GdFeCo bilayer (open markers). The experimental data is scaled so that it agrees with the thermal model predictions at 300 ps (solid and dashed lines). The value for $g_{ep,GFC}$ was derived by fitting the model predictions to the data at delay times less than 5 picoseconds.

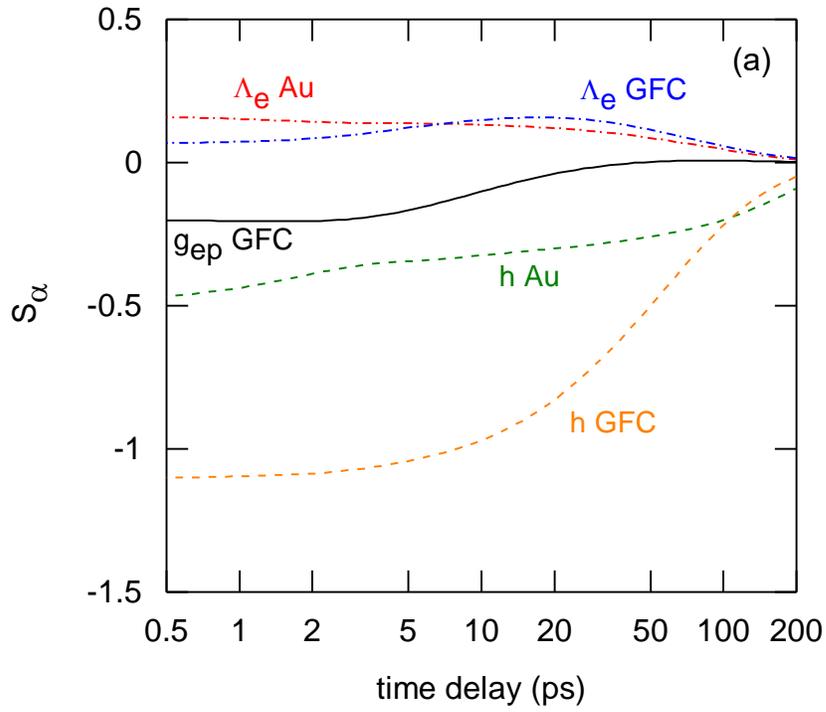

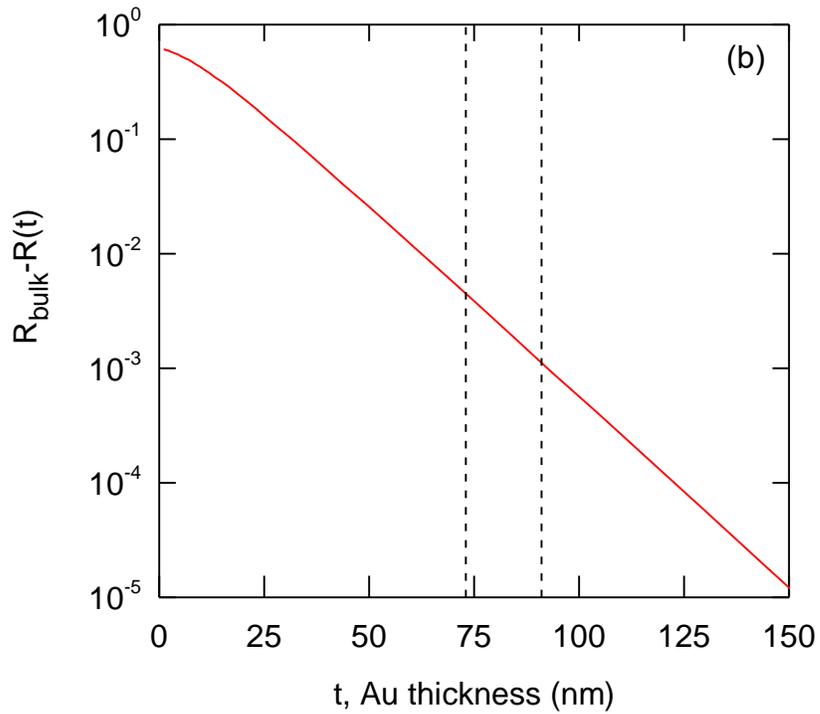

**Figure 3.** (a) Sensitivity coefficients for temperature of the Au phonons for the 73 nm Au / 11.5 nm GdFeCo bilayer. The phonon temperature is most sensitive to the thicknesses of the Au and GdFeCo layers, therefore 5 and 10% uncertainties in these layers are responsible for most of our experimental uncertainty in $g_{ep,GFC}$. (b) Predicted difference between the reflectance of a Au/GFC bilayer and bulk Au as a function of the Au layer thickness. In order to guarantee the GdFeCo temperature evolution did not contribute to the experimental data in Fig. 2, an optically thick layer of Au is required. Dashed vertical lines indicate the Au thicknesses of the samples considered in this study.